\documentclass[twocolumn,trackchanges]{aastex631}

\usepackage{CJK}
\turnoffeditone

\newcommand{\petit}{\texttt{petitRADTRANS}}

\newcommand{\dynesty}{\texttt{dynesty}}
\newcommand{\teff}{T$_{\rm eff}$}

\newcommand{\kms}{$\rm km\,s^{-1}$ }
\newcommand{\caltech}{Department of Astronomy, California Institute of Technology, Pasadena, CA 91125, USA}
\newcommand{\gps}{Division of Geological \& Planetary Sciences, California Institute of Technology, Pasadena, CA 91125, USA}
\newcommand{\ucsc}{Department of Astronomy \& Astrophysics, University of California, Santa Cruz, CA95064, USA}
\newcommand{\keck}{W. M. Keck Observatory, 65-1120 Mamalahoa Hwy, Kamuela, HI 96743, USA}
\newcommand{\ucla}{Department of Physics \& Astronomy, 430 Portola Plaza, University of California, Los Angeles, CA 90095, USA}
\newcommand{\jpl}{Jet Propulsion Laboratory, California Institute of Technology, 4800 Oak Grove Dr.,Pasadena, CA 91109, USA}
\newcommand{\ucsd}{Center for Astrophysics and Space Sciences, University of California, San Diego, La Jolla, CA 92093}

\newcommand{\northwestern}{Center for Interdisciplinary Exploration and Research in Astrophysics (CIERA) and Department of Physics and Astronomy,
Northwestern University, Evanston, IL 60208, USA}
\newcommand{\arizona}{James C. Wyant College of Optical Sciences, University of Arizona,
Meinel Building 1630 E. University Blvd., Tucson, AZ 85721, USA}

\received{September 15, 2023}
\revised{November 18, 2023}
\accepted{November 28, 2023}

\shorttitle{Metallicity and C/O of HD 189733 b}
\shortauthors{Finnerty et al.}
\graphicspath{{./}{}}

\begin{document}
\begin{CJK*}{UTF8}{gbsn}

\title{Atmospheric metallicity and C/O of HD 189733 b from high-resolution spectroscopy}

\correspondingauthor{Luke Finnerty}
\email{lfinnerty@astro.ucla.edu}

\author[0000-0002-1392-0768]{Luke Finnerty}
\affiliation{\ucla}

\author[0000-0002-6618-1137]{Jerry W. Xuan}
\affiliation{\caltech}

\author[0000-0002-6171-9081]{Yinzi Xin}
\affiliation{\caltech}

\author[0000-0002-4934-3042]{Joshua Liberman}
\affiliation{\caltech}
\affiliation{\arizona}

\author{Tobias Schofield}
\affiliation{\caltech}

\author[0000-0002-0176-8973]{Michael P. Fitzgerald}
\affiliation{\ucla}

\author[0000-0003-2429-5811]{Shubh Agrawal}
\affiliation{Department of Physics and Astronomy, University of Pennsylvania, Philadelphia, PA 19104, USA}

\author{Ashley Baker}
\affiliation{\caltech}

\author{Randall Bartos}
\affiliation{\jpl}

\author{Geoffrey A. Blake}
\affiliation{\gps}


\author[0000-0003-4737-5486]{Benjamin Calvin}
\affiliation{\caltech}
\affiliation{\ucla}

\author{Sylvain Cetre}
\affiliation{\keck}

\author[0000-0001-8953-1008]{Jacques-Robert Delorme}
\affiliation{\keck}
\affiliation{\caltech}

\author{Greg Doppmann}
\affiliation{\keck}

\author{Daniel Echeverri}
\affiliation{\caltech}

\author[0000-0002-5370-7494]{Chih-Chun Hsu}
\affiliation{\northwestern}

\author[0000-0001-5213-6207]{Nemanja Jovanovic}
\affiliation{\caltech}

\author[0000-0002-2019-4995]{Ronald A. L\'opez}
\affiliation{\ucla}

\author[0000-0002-0618-5128]{Emily C. Martin}
\affiliation{\ucsc}

\author{Dimitri Mawet}
\affiliation{\caltech}
\affiliation{\jpl}

\author{Evan Morris}
\affiliation{\ucsc}

\author{Jacklyn Pezzato}
\affiliation{\caltech}

\author[0000-0003-2233-4821]{Jean-Baptiste Ruffio}
\affiliation{\ucsd}

\author[0000-0003-1399-3593]{Ben Sappey}
\affiliation{\ucsd}

\author{Andrew Skemer}
\affiliation{\ucsc}

\author{Taylor Venenciano}
\affiliation{Physics and Astronomy Department, Pomona College, 333 N. College Way, Claremont, CA 91711, USA}

\author[0000-0001-5299-6899]{J. Kent Wallace}
\affiliation{\jpl}

\author[0000-0003-0354-0187]{Nicole L. Wallack}
\affiliation{Earth and Planets Laboratory, Carnegie Institution for Science, Washington, DC 20015, USA}

\author[0000-0003-0774-6502]{Jason J. Wang (王劲飞)}
\affiliation{\northwestern}

\author[0000-0002-4361-8885]{Ji Wang (王吉)}
\affiliation{Department of Astronomy, The Ohio State University, 100 W 18th Ave, Columbus, OH 43210 USA}



\begin{abstract}
We present high-resolution $K$-band emission spectra of the quintessential hot Jupiter HD 189733 b from the Keck Planet Imager and Characterizer (KPIC). Using a Bayesian retrieval framework, we fit the dayside pressure-temperature profile, orbital kinematics, mass-mixing ratios of H$_2$O, CO, CH$_4$, NH$_3$, HCN, and H$_2$S, and the $\rm ^{13}CO/^{12}CO$ ratio. We measure mass fractions of $\rm \log H_2O = -2.0^{+0.4}_{-0.4}$ and $\rm \log CO = -2.2^{+0.5}_{-0.5}$, and place upper limits on the remaining species.  Notably, we find $\rm \log CH_4 < -4.5$ at 99\% confidence, despite its anticipated presence at the equilibrium temperature of HD 189733 b assuming local thermal equilibrium. We make a tentative ($\sim3\sigma$) detection of $\rm ^{13}CO$, and the retrieved posteriors suggest a $\rm ^{12}C/^{13}C$ ratio similar to or substantially less than the local interstellar value. The possible $\rm ^{13}C$ enrichment would be consistent with accretion of fractionated material in ices or in the protoplanetary disk midplane. The retrieved abundances correspond to a substantially sub-stellar atmospheric $\rm C/O = 0.3\pm0.1$, while the carbon and oxygen abundances are stellar to slightly super-stellar, consistent with core-accretion models which predict an inverse correlation between C/O and metallicity. The specific combination of low C/O and high metallicity suggests significant accretion of solid material may have occurred late in the formation process of HD 189733 b. 

\end{abstract}



\section{Introduction} \label{sec:intro}

First discovered in 2005 through both transit and radial velocity observations \citep{bouchy2005}, the hot Jupiter HD 189733 b is a frequent target for characterization studies due to its bright K-type host star ($K_{\rm mag} = 5.5$, \citealt{cutri2003}) and large transit depth ($\sim2\%$). \citet{winn2006} measured the Rossiter-McLaughlin effect for the system, finding the planetary orbit to be well aligned with the stellar rotation axis ($\lambda = -1.4^{\circ}\pm1.1^{\circ}$). \citet{deming2006} detected the secondary eclipse in \textit{Spitzer} observations, eventually leading to the first full-disk temperature map for an exoplanet \citep{knutson2007} which found the hottest part of the planet is offset by $16^{\circ}\pm6^{\circ}$ east from the substellar point due to supersonic winds.

Early efforts to characterize the chemical composition of HD 189733 b were met with mixed results, a selection of which we summarize here. Some studies reported detection of H$_2$O \citep{tinetti2007, grillmair2008} or CO \citep{desert2009}, while others reported a flat transmission spectrum \citep{grillmair2007, pont2008, sing2009}. The flat spectra were found to be consistent with the presence of hazes \citep{pont2008, sing2009}, preventing the clear detection of individual molecular species from low-resolution transmission spectra. Subsequent observations suggest this haze extends from the optical into the near-infrared $H$ band \citep{gibson2012}.

High-resolution emission spectroscopy provides a potential pathway to perform atmospheric characterization in the presence of clouds or hazes. Since high-resolution spectroscopy probes line cores at lower pressures compared to transmission spectroscopy, emission originating above the cloud layer can be detected and characterized \citep{gandhi2020}. \citet{dekok2013} used high-resolution transmission spectroscopy to detect CO emission and place an upper limit on the H$_2$O line contrast. \citet{brogi2016} used transmission spectroscopy to measure a rotation rate consistent with tidal locking and a blueshift suggesting a $\sim$2\,\kms day-to-night wind. Water was eventually detected in high-resolution transmission spectroscopy from GIANO \citep{brogi2018} and has been repeatedly confirmed in both emission and transmission spectroscopy \citep{flowers2019, cabot2019, brogi2019, Boucher2021, klein2023}. CO has similarly been confirmed in multiple studies \citep{brogi2019, cabot2019, flowers2019}, while \citet{cabot2019} also reported tentative evidence for HCN. 

Despite the repeated confirmation of the presence of both H$_2$O and CO, none of these studies were able to obtain simultaneously bounded constraints on the abundances of both species. This has prevented robust measurement of the C/O ratio and bulk atmospheric metallicity, potentially key diagnostics to trace the formation and evolution of the HD 189733 system. Additionally, despite the anticipated presence of CH$_4$ under local thermal equilibrium at the equilibrium temperature of HD 189733 b, none of these studies made a clear CH$_4$ detection, leaving an important potential tracer of photochemistry effects unconstrained. 

To address these gaps in our knowledge of an important benchmark hot Jupiter system, we observed HD 189733 b with Keck/KPIC high-resolution $K$-band spectroscopy. These observations are part of an ongoing survey to characterize a statistically significant number of hot Jupiter atmospheres with KPIC and constrain the underlying distribution of C/O and metallicity in this population.

Section \ref{sec:obs} describes the observations, data reduction procedures, and atmospheric retrieval framework, with an emphasis on the differences from \citet{finnerty2023}. Section \ref{sec:res} presents the results of the atmospheric retrievals and cross-correlation analysis. Section \ref{sec:disc} discusses the retrieved atmospheric properties, comparing the measured abundances with expectations from chemical equilibrium and photochemical models. We summarize our results and draw conclusions in Section \ref{sec:conc}.

\section{Observations and Data Reduction}\label{sec:obs}

\begin{deluxetable}{ccc}
    \tablehead{\colhead{Property} & \colhead{Value} & \colhead{Ref.}}
    \startdata
        & \textbf{HD 189733} & \\
        \hline
        RA & 20:00:43.7 & \citet{gaiaedr3} \\
        Dec & +22:24:39.1 & \citet{gaiaedr3} \\
        Spectral Type & K2V &  \\
        $K_{\rm mag}$ & $5.54\pm0.02$ & \citet{cutri2003} \\
        Mass & $0.82\pm0.03$ $\rm M_\odot$ & \citet{rosenthal2021}  \\
        Radius & $0.78\pm0.01$ $\rm R_\odot$ & \citet{rosenthal2021} \\
        \teff & $5005\pm77$ K & \citet{polanski2022} \\
        $\log g$ & $4.49\pm0.09$ & \citet{polanski2022} \\
        $v\sin i$ & $1.8\pm0.9$  \kms & \citet{polanski2022}\\
        $v_{\rm rad}$ & $-2.3\pm0.2$ \kms & \citet{gaiarv2}  \\
        $\rm [Fe/H]$ & $0.01\pm0.03$ & \citet{polanski2022}  \\
        $\rm [C/H]$ & $-0.12\pm0.05$ & \citet{polanski2022} \\
        $\rm [O/H]$ & $ -0.21\pm0.07$ & \citet{polanski2022} \\
        C/O & $0.67\pm0.14$ & \citet{polanski2022} \\
        \smallskip \\
         & \textbf{HD 189733 b} & \\
        \hline
        Period & 2.21857567 days  & \citet{exofop3}  \\
        $\rm t_{\rm transit}$ & JD 2454279.436714 & \citet{exofop3}  \\
        $a$ & 0.031 AU & \citet{rosenthal2021}  \\
        $i$ & 85.7$^\circ$ & \citet{stassun2017} \\
        $K_{\rm p}$ & 153 \kms & Estimated  \\
        Mass & $1.13\pm0.08$ $\rm M_{Jup}$ & \citet{stassun2017} \\
        Radius & $1.13\pm0.01$ $\rm R_{Jup}$ & \citet{stassun2017} \\
        $\rm T_{\rm eq}$ & 1200 K & Estimated \\
        C/O & $0.3\pm0.1$ & This work  \\
        $\rm [C/H]$ & $+0.4\pm0.5$ & This work \\
        $\rm [O/H]$ & $+0.8\pm0.4$ & This work
    \enddata
    \caption{Stellar and planetary properties for the HD 189733 system.} 
    \label{tab:props}
\end{deluxetable}

\subsection{Observations}

HD 189733 was observed with Keck/KPIC phase II \citep{kpic, echeverri2022} on 2022 August 13 (UT) from 8:39 UT to 11:59 UT at an airmass of 1.0 to 1.36. KPIC provides a single-mode fiber feed into Keck/NIRSPEC \citep{nirspec, nirspecupgrade, nirspecupgrade2}, enabling stable diffraction-limited high-resolution spectroscopy. The observations began approximately 30 minutes after the expected end of secondary eclipse, covering orbital phases from 0.5285 to 0.5898, during which time we expect the star/planet relative radial velocity to shift from $-27$ \kms to $-82$ \kms. Observations were taken with an ABBA nodding pattern between KPIC science fibers 2 and 4 with 30 second exposures, giving a total of 256 science frames. 

The KPIC real-time throughput calculator (included in the KPIC DRP) estimated the 95th-percentile top-of-atmosphere throughput was 3--3.4\% at the start of the observations, decreasing to $\sim2.5\%$ in the second half of the sequence, roughly consistent with the typical performance measured for KPIC phase II under photometric conditions \citep{echeverri2022}. Weather conditions were clear and stable throughout the observations. The throughput loss is likely a result of an increase in the PSF size due to seeing and AO performance changes as the airmass increased from 1.0 at the start of the observation sequence to 1.36 by the end. The median extracted per-pixel signal-to-noise ratio after coadding frames in a nodding sequence was approximately 180, ranging from $\sim210$ in the bluest order to $\sim160$ in the reddest order. 

\subsection{Data Reduction}

\begin{figure*}
    \centering
    \includegraphics[width=1.0\linewidth]{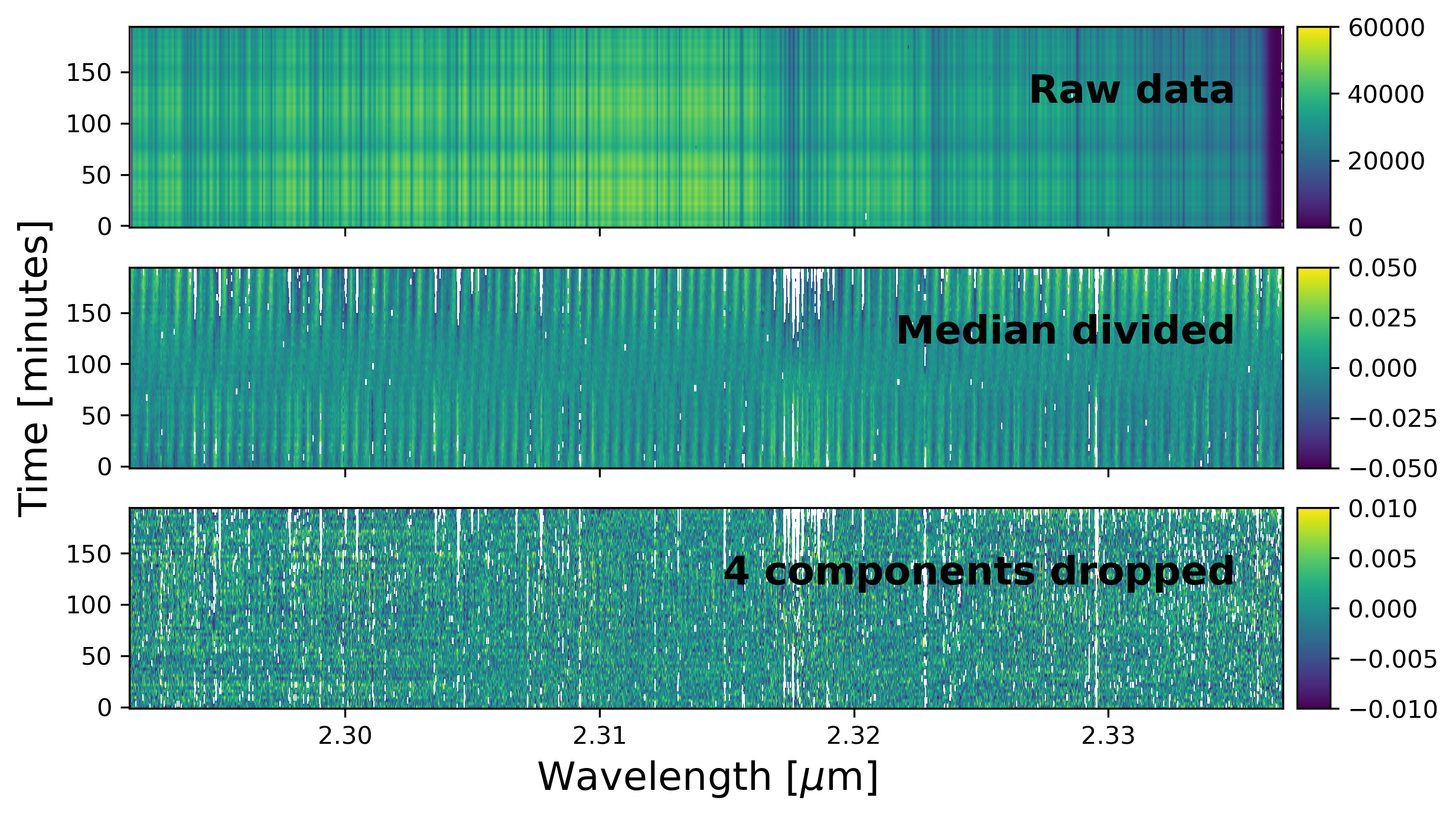}
    \caption{Data detrending process for KPIC HD 189733 observations. The top row shows the raw time series, with significant frame-to-frame flux variations, tellurics, and blaze function effects. The colorbar for the top frame is the counts in the extracted 1D spectrum. The second row shows the same data after scaling each exposure by its median and dividing each frame by the time-series median, with the colorbar indicating the median-normalized and continuum-subtracted flux level. The blaze and telluric signals are almost completely eliminated, but a significant time-varying fringe is clearly present. White pixels indicate data that have been masked. In the bottom panel, the fringe has been effectively removed by dropping the first 4 components of the SVD decomposition of the time series. The bottom colorbar shows the median-normalized, continuum-subtracted flux level after the decomposition.}
    \label{fig:reduction}
\end{figure*}

The data extraction is identical to that described in \citet{finnerty2023}, which differs slightly from the current version of the supported KPIC Data Reduction Pipeline (DRP)\footnote{\href{https://github.com/kpicteam/kpic_pipeline/}{https://github.com/kpicteam/kpic\_pipeline/}}. HIP 95771 (spectral type M0.5IIIb) was used for wavelength calibration at the start of the night. Of the nine observed NIRSPEC orders, three (orders 37-39) are heavily contaminated by telluric CO$_2$ lines, while two (orders 35 and 36) are almost entirely lacking in stellar or telluric lines, preventing an accurate wavelength calibration. We therefore use the four reddest orders (orders 31-34) spanning $\sim$2.2--2.5\,\micron\ (with significant gaps) for which we have accurate wavelength solutions and minimal telluric contamination. 

In order to speed the log-likelihood calculation in the retrieval, we coadded the extracted spectra from each ABBA sequence after re-interpolating all observations onto the science fiber 2 wavelength solution. Each fiber has a slightly different line-spread function (LSF) and the fiber coupling can vary significantly between frames depending on the AO correction, leading to variability in the coadded LSF. The final time series consists of 64 spectra, with a maximum planetary radial velocity shift between {\em consecutive} spectra of approximately 900 $\rm m\,s^{-1}$, roughly 10$\times$ smaller than the NIRSPEC resolution ($\rm R\sim35,000$, $\rm \Delta v\sim8.5$ \kms for the chosen instrument settings). 

The data detrending process for one order is shown in Figure \ref{fig:reduction}. The extracted spectral time-series for each order is first scaled to have a median of 1.0, then divided by the time-series median spectrum using a second-order polynomial to fit continuum variations. This removes most temporally fixed features in the spectrum, leaving time-varying fringing \citep[][Xuan et al. submitted]{Finnerty2022, finnerty2023spie}, airmass-driven telluric variations, and the time-varying planet signal. Points differing from the median by more than $6\times$ the median absolute deviation (MAD) are then masked, as well as the first and last fifty wavelength bins in each order. 

The time series for each order is then mean-subtracted and the first 4 components of the singular value decomposition (SVD) are projected out. To account for the resulting distortion of the planet signal, the dropped components are saved and added to the forward model during the log-likelihood calculation. A new SVD is then performed on the component-injected forward model in order to replicate the distortion of the planet signal from projecting these components out, similar to \citet{line2021}. In contrast with \citet{finnerty2023}, we do not negatively inject the forward model prior to removal of the SVD components, which significantly speeds the log-likelihood calculation. This choice makes a stronger assumption that repeating the SVD on the component-injected model accurately reproduces the distortion of the observed planet signal, but has proven successful in previous work \cite[e.g.][]{line2021}. After the decomposition, points varying by more than $4\times$ the MAD are masked. Previously, \citet{finnerty2023} found several orders of KPIC observations continued to show a fringing pattern that dominated over the Gaussian noise even at this stage of the analysis, which was removed by taking a final median division. In these observations, we do not see any evidence of similar residual fringing above the noise level (for example in the third panel of Figure \ref{fig:reduction}), and therefore we omit this step.

\subsection{Atmospheric Retrieval}

The atmospheric retrieval framework is similar to that described in \citet{finnerty2023}, though we have implemented several changes to improve the time-series detrending. We briefly summarize this framework for completeness. The fit parameters, priors, and retrieved values are listed in Table \ref{tab:priors}.

As in \citet{finnerty2023}, we use \petit\ \citep{prt:2019, prt:2020} for the radiative-transfer model of the planetary atmosphere with 80 log-uniform spaced pressure layers between $10^2$ and $10^{-6}$ bar. We use the 4-parameter pressure-temperature ($P-T$) profile model from \citet{guillot2010}, implemented in \petit. From initial tests using the \citet{madhusudhan2009} 6-parameter $P-T$ model previously used in \citet{finnerty2023}, we found that multiple parameters were poorly constrained and showed complex degeneracies. The \citet{guillot2010} parameterization is more constraining in its physical assumptions, making it better-suited to the limited wavelength range of these observations and enabling looser, physically motivated priors for the $P-T$ parameters. Retrieved molecular abundances were generally consistent between both parameterizations. At all pressures, we require $\rm 100\, K < T < 3000\, K$ for physical consistency and to avoid hitting the bounds of our opacity tables. We also include a grey cloud deck and use the scattering mode of \petit. 

We fit for vertically-fixed abundances (not necessarily in chemical equilibrium) of H$_2$O, $^{12}$CO, CH$_4$, NH$_3$, HCN, H$_2$S, and the $^{13}$CO/$^{12}$CO ratio. For H$_2$O, $^{12}$CO, and $^{13}$CO we used the high-temperature opacity tables described in \citet{finnerty2023}. For CH$_4$, we used the opacity table computed in \citet{xuan2022} from the \citet{hargreaves2020} linelist. For NH$_3$ \citep[linelist:][]{yurchenko2011}, HCN \citep[linelist:][]{harris2006, barber2014}, and H$_2$S \citep[linelist:][]{rothman2013}, we used the \petit\ high-resolution opacity tables described in \citet{molliere2019}. Collision induced absorption (CIA) from $\rm H_2 - H_2$ and $\rm H_2 - He$ are also included. We fit for the abundance of H$_2$, though we do not include H$_2$ line opacity. The remaining mass of the atmosphere is assumed to be helium. As we do not include H$_2$ line opacity, the retrieved H$_2$ abundance reflects the best-fit mean-molecular weight via the impact on the CIA, rather than strengths of H$_2$ lines in the planetary spectrum. We use a PHOENIX model for the star with $T_{\text{eff}} = 5000 \rm\,K$, $\log g = 4.5$, and $\rm [Fe/H] = 0.0$, which has been broadened to $v\sin i = 1.8$ \kms \citep{polanski2022}. While \citet{finnerty2023} fixed the size of the Gaussian broadening kernel to a value found by varying the kernel size for a fixed planet model and maximizing the likelihood, in this work we set the kernel width to $\sigma = 1.2$ pixels, slightly smaller than the expected value for the instrument, and apply a rotational broadening kernel to the atmospheric forward model using the fast technique described in \citet{carvalho2023} which accounts for the wavelength-dependence of the rotational kernel. The size of this kernel is a free parameter in the fit, allowing us to determine the best-fit line width accounting for coadd-induced LSF variations and the change in projected planetary velocity within a single coadd. 

The nested sampling was performed with \texttt{dynesty} \citep{speagle2020} using 1200 live points, running to a $\Delta \log z = 0.01$ stopping criteria. The retrieval took approximately four days with 16 Intel E5-2670 CPU cores, totaling about 60 days of CPU time. The significant increase in runtime compared with \citet{finnerty2023} is a result of including scattering in the radiative transfer calculation.

\section{Results}\label{sec:res}

\begin{deluxetable*}{ccccc}
    \tablehead{\colhead{Name} & Symbol  & \colhead{Prior} & \colhead{Retrieved Max-L} & \colhead{Retrieved Median}}
    \startdata
        log infrared opacity [$\rm cm^{2} g^{-1}$] & $\log \kappa$ &  Uniform(-3, 0) & -1.0 & $-1.1^{+0.4}_{-0.4}$ \\
       log infrared/optical opacity & $\log \gamma$  &  Uniform(-1.5,-0.3) & -1.4 & $-1.3^{+0.3}_{-0.2}$ \\
        Internal temperature [K] & $\rm T_{int}$ & Uniform(50, 300) & 140 & $200^{+100}_{-100}$ \\
        Equilibrium temperature [K] & $\rm T_{equ}$ & Uniform(400,1500) & 800 & $800^{+100}_{-100}$ \\
        Grey cloud pressure [bar] & $\rm \log P_{cloud}$ & Uniform(-2,1) & -1.0 & $-0.3^{+0.9}_{-0.6}$ \\
        $K_{\rm p}$ offset [\kms] & $\Delta K_{\rm p}$  & Uniform(-40, 40) &  6.4 & $7.0^{+7.7}_{-8.3}$\\
        $v_{\rm sys}$ offset [\kms] & $\Delta v_{\rm sys}$ & Uniform(-15, 15) & -7.3 & $-6.4^{+3.4}_{-3.6}$ \\
        Velocity broadening kernel [\kms] & $v_{\rm broad}$ & Uniform(1,20) & 16.9 & $16.0^{+1.0}_{-1.0}$ \\
        log H$_2$O mass-mixing ratio & log H$_2$O  &  Uniform(-4, -0.3) & -1.7 & $-2.0^{+0.4}_{-0.4}$ \\
        log CO mass-mixing ratio & log CO &  Uniform(-4, -0.3) & -1.8 & $-2.2^{+0.5}_{-0.5}$ \\
        log CH$_4$ mass-mixing ratio & log CH$_4$  & Uniform(-8,-1) & -7.4 & $-6.7^{+1.0}_{-0.9}$ \\
        log NH$_3$ mass-mixing ratio & log NH$_3$  & Uniform(-8,-2) & -4.8 & $-6.5^{+1.1}_{-1.0}$\\
        log HCN mass-mixing ratio & log HCN  & Uniform(-8,-2) & -5.8 & $-5.4^{+1.4}_{-1.7}$ \\
        log H$_2$S mass-mixing ratio & log H$_2$S  & Uniform(-8,-3) & -7.7 & $-5.6^{+1.7}_{-1.6}$\\
        log $\rm ^{13}CO/^{12}CO$ & $\rm log ^{13}CO_{rat}$ &  Uniform(-3, -0.3) & -0.8 & $-1.0^{+0.4}_{-0.8}$ \\
        log H$_2$ mass-mixing ratio & log H$_2$ &  Uniform(-0.4, -0.05) & -0.36 & $-0.26^{+0.13}_{-0.10}$ \\
        Scale factor & scale & Uniform(0,5) & 4.9 & 4.5$^{+0.4}_{-0.7}$  
        \smallskip \\
         & & \textbf{Derived Parameters} & \\
        \hline
        Carbon/oxygen ratio & C/O & - & 0.32 & $0.3\pm0.1$ \\
        Carbon abundance & [C/H] & - & $-2.6$ & $-3.1\pm0.5$ \\
        Oxygen abundance & [O/H] & - & $-2.1$ & $-2.5\pm0.4$ \\
    \enddata 
    \caption{List of parameters, priors, and results for atmospheric retrievals. The error bars on the retrieved medians correspond to the 68$\% / 1\sigma$ confidence interval. In addition to these priors, we required both a non-inverted $P-T$ profile and that the atmospheric temperature stay below 3000 K at all pressure levels. The full corner plot is included in Appendix \ref{app:corner}.}
    \label{tab:priors}
\end{deluxetable*}

Table \ref{tab:priors} presents the priors and results from the retrieval, including both the maximum-likelihood and median retrieved values. We discuss the velocity parameters in Section \ref{ssec:wind}, the $P-T$ profile and thermal properties in Section \ref{ssec:thermal}, and the chemical abundances in Section \ref{ssec:chem}. The full corner plot is included in Appendix \ref{app:corner}.

\begin{figure*}
    \centering
    \includegraphics[width=0.9\linewidth]{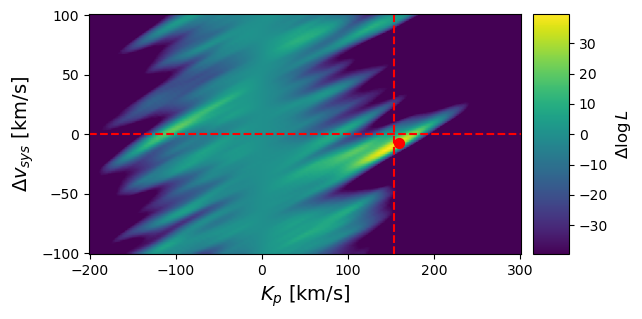} 
    \caption{$v_{\rm sys} - K_{\rm p}$ diagram for the maximum-likelihood model. The planet is clearly detected, though with a significant blueshift compared to the expectation indicated in dashed red. The detrending process suppresses planet features at small $K_{\rm p}$, producing the feature centered on $K_{\rm p} = 0$, where the planet model is a flat line after detrending. Compared with this null case, the maximum-likelihood model is preferred by $\Delta \log L = 39$, equivalent to $\sim6\sigma$ using Wilks' Theorem \citep{wilks1938} with 17 free parameters. Note that the significant off-peak structure prevents accurate estimation of detection significance from division by the standard deviation far from the planet feature.}
    \label{fig:kpvsys}
\end{figure*}

\subsection{Velocity and winds}\label{ssec:wind}

The $v_{\rm sys} - K_{\rm p}$ diagram for the maximum-likelihood planet model is shown in Figure \ref{fig:kpvsys}. Note that self-division of the planet spectrum results in substantial structure far from the planet peak, which prevents an accurate estimation of detection significance from dividing by the off-peak standard deviation. This effect can also be seen in Figure 6 of \citet{finnerty2023}, and is a general feature of detrending schemes which rely on the velocity shift of planet features that is then compounded by the explicit dependence of the \citet{brogi2019} log-likelihood function on the forward-model variance. As an alternative to estimate the detection strength, we use Wilks' Theorem \citep{wilks1938} with 17 free parameters to estimate the significance of the detection compared with a flat planet model to be $\sim6\sigma$. Alternatively, we note that the off-peak structure is a strong function of $K_{\rm p}$. Subtracting each column of the $v_{\rm sys} - K_{\rm p}$ by its median, while not statistically robust, mostly removes this structure and allows a more reliable estimate of the off-peak variance. Using the $K_{\rm p} < 0$ values to estimate the standard deviation after this subtraction gives a detection strength of $7.8\sigma$. More rigorously, the cross-correlation coefficient depends only indirectly on the model variance, and the variance of the cross-correlation coefficient is therefore a weaker function of $K_{\rm p}$, though self-division of the model still has some impact at small $K_{\rm p}$. Computing the $v_{\rm sys} - K_{\rm p}$ diagram using the cross-correlation coefficient and dividing by the standard deviation of the $K_{\rm p} < 0$ region gives a detection strength of $7.3\sigma$ near the retrieved maximum-likelihood velocity parameters. 

The retrieved planet velocity is blueshifted by several \kms. Based on the orbital phase sampling of the observations, we expect that the previously reported day-to-night wind would appear as a redshift. The apparent blueshift is still consistent with values in the literature \citep[e.g.][]{Boucher2021, klein2023}, but may suggest a more complicated circulation pattern than a single day-night wind. Alternatively, a number of other factors could bias our retrieved velocity, which we discuss in the next section.

The broadening velocity reported in Table \ref{tab:priors} is substantially larger than values previously reported for the planetary rotational velocity in the literature \citep[e.g. $3.4^{+1.3}_{-2.1}$ \kms in][]{brogi2016} and is inconsistent with the expectation that HD 189733 b be tidally locked, which would correspond to a rotational velocity of 2.6 \kms. This is due to several simplifying assumptions in our handling of the KPIC LSF which are absorbed into a larger broadening velocity. Specifically, we assumed a smaller nominal LSF smaller than \citet{finnerty2023}, allowing the retrieval to freely determine the best-fit line width. This helps account for LSF variations between coadded science spectra that result from the slight differences in LSFs between science fibers and the varying frame-to-frame coupling efficiency, and can also help account for the line smearing due to the planet's motion within a coadd. While this prevents us from actually measuring the physical rotational speed of HD 189733 b, the tightly constrained posterior suggests these data have sufficient sensitivity to the planetary line shapes that future analyses will be able to measure $v\sin i$ for hot Jupiters.

\subsection{Thermal properties}\label{ssec:thermal}

\begin{figure*}
    \centering
    \includegraphics[width=1.0\linewidth]{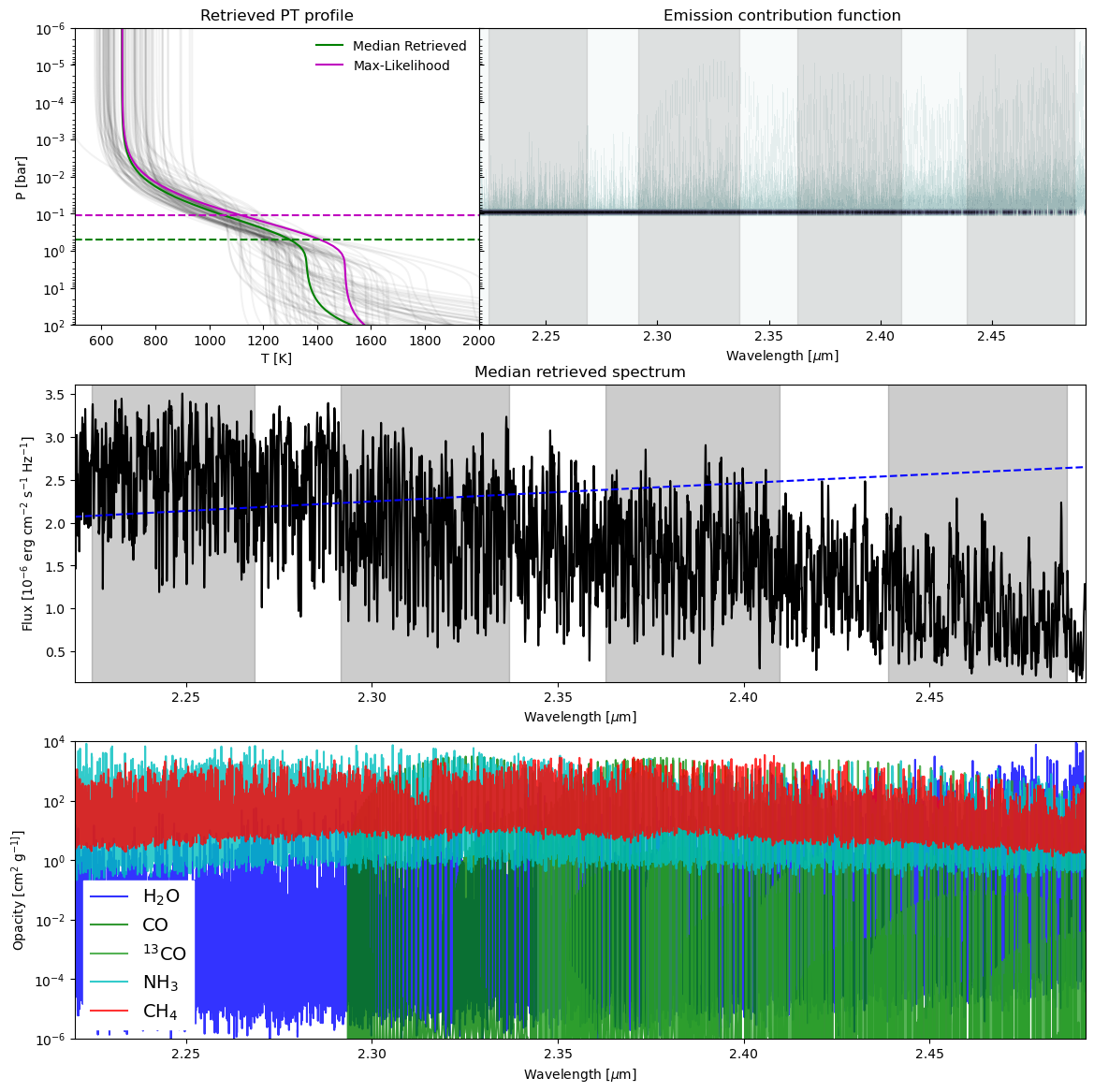}
    \caption{Retrieved $P-T$ profile (top left), maximum-likelihood emission contribution function (top right), maximum-likelihood planet spectrum (middle), and opacities for H$_2$O, CO, NH$_3$, and CH$_4$ (bottom). The observed NIRSPEC orders are shaded in grey. In addition to the maximum-likelihood and median $P-T$ profiles, the top left also includes the corresponding cloud top pressures as dashed horizontal lines and the $P-T$ profiles from 100 draws from the retrieved posterior. While several parameters are poorly constrained in the corner plots, the actual $P-T$-profiles follow a tight distribution. The emission contribution function shows the emission mostly arises near 100 mbar, just above the cloud deck, with contribution from higher altitudes in the CO line cores. The dashed blue line plotted with the maximum-likelihood spectrum shows the flux from a 1200 K blackbody, which as expected is comparable to the retrieved planet flux. The slope of the retrieved spectrum differs from a blackbody as a result of CO and especially H$_2$O absorption features at the red end of the $K$-band. Our retrievals provide only upper limits on NH$_3$ and CH$_4$ abundances despite these species' substantial  opacity across the entire observed band, suggesting that the limits indicate a real absence of these species.}
    \label{fig:spec}
\end{figure*}

Figure \ref{fig:spec} shows the retrieved $P-T$ profile, emission contribution function, maximum-likelihood spectrum, and opacities of the major species. While several of the $P-T$ parameters are poorly constrained in the 1D marginalized posteriors, repeated draws from the posterior yield a consistent set of profiles, suggesting the $P-T$ profile is over-parameterized for the available data, but is well-constrained in the physically significant space.  

The retrieved $P-T$ profile is slightly cooler than previous theoretical models for HD 189733 b \citep[e.g.][]{vulcan}. This may contribute to the preference for larger values of the overall multiplicative scale factor applied to the model planet spectrum, which results in an overall flux level consistent with the expected equilibrium temperature for HD 189733 b (Figure \ref{fig:spec}, middle panel). Alternatively, lower cloud pressures could be counteracted by increasing the scale factor in order to maintain the effective strength of the planet lines relative to the stellar continuum. Additionally, \citet{finnerty2023} found in simulations that this retrieval framework prefers larger values of the scale factor, likely due to artifacts in the data processing.

Due to this uncertainty, we ran an additional retrieval with a Uniform(0,10) prior on the scale factor. The scale factor was again close to the upper bound, and the retrieved $P-T$ profile was even colder. However, the retrieved abundances were consistent at the $1\sigma$ level and the flux level of the maximum-likelihood planet model was consistent to within $\sim10\%$ for both retrievals. Initial analysis of another hot Jupiter observed with KPIC (Finnerty et al. in prep.) is also showing a similar preference for a higher-than-expected scale factor that is countered by a cooler-than-expected $P-T$ profile to match the expected continuum level. This strongly suggests our free retrieval framework has limited sensitivity to absolute temperature from the $K$-band data alone, but that this uncertainty does not significantly impact the retrieved atmospheric composition. Finally, we also note that residual continuum slopes or offsets in either the data or the model could also result in a spurious preference for a larger scale factor. In this case, the scale factor parameter is attempting to replicate the continuum offset, while the $P-T$ profile is shifted in order to preserve the correct line strength relative to the scaled continuum.

\subsection{Chemical composition}\label{ssec:chem}

Of the included species, we obtain bounded constraints only for CO and H$_2$O. Of the remaining species, we obtain upper limits on CH$_4$, NH$_3$, and HCN, though we note that the marginalized HCN posterior shows weak preference for $\rm \log HCN \sim -4.8$. These species all have opacities comparable to that of the 2.3 $\mu$m CO bandhead in at least one of the observed NIRSPEC orders, suggesting these species would have been detected if present in significant abundances. While H$_2$S is effectively unconstrained, the opacity table used has a gap from $2.25-2.35\rm\ \mu m$, which may preclude detection. The retrieved posterior also prefers a very high $\rm ^{13}CO/^{12}CO$ ratio, peaking at $\sim 10^{-0.8}$, though with a substantial tail to lower values. We include detection maps for H$_2$O, CO, CH$_4$, and $\rm ^{13}CO$ in Appendix \ref{app:detmaps}, as well as a discussion of the challenges associated with accurately estimating single-molecule detection strength.

\section{Discussion}\label{sec:disc}

\subsection{Comparison to previous results}

The extensive existing literature on HD 189733 b provides a basis for comparison with our retrieval results. To more easily facilitate comparison to other results, we note that the median retrieved values values reported in Table \ref{tab:priors} are equivalent to $K_{\rm p} = 160\pm8$ \kms, $\rm \log CO_{VMR} = -3.3\pm0.5$, and $\rm \log H_2O_{VMR} = -2.9\pm0.4$. The remainder of this paper will continue to use mass-mixing ratios for molecular abundances. 

\citet{klein2023} analyzed high-resolution infrared transmission spectra from CFHT/SPIROU \citep{donati2020} to perform an atmospheric retrieval of the H$_2$O abundance and $P-T$ profile using the same \citet{guillot2010} parameterization with a grey cloud deck. \citet{klein2023} report $K_{\rm p}$, $v_{\rm sys}$, and $\rm T_{eq}$ in good agreement with the values in Table \ref{tab:priors}, and $\rm \log H_2O = -2.95^{+0.75}_{-0.53}$, slightly smaller than the value in Table \ref{tab:priors}, while reporting a larger value for the cloud-top pressure $\rm \log P_{cloud} = 0.479^{+1.02}_{-1.06}$. We note that the corner plots for both retrievals indicate a degeneracy between the cloud pressure and molecular abundances, with greater pressures corresponding to lower abundances, potentially explaining the minor discrepancy in these parameters.

The data analyzed in \citet{klein2023} were originally described in \citet{Boucher2021}, who reported a somewhat lower H$_2$O abundance, higher cloud pressure, and lower temperature. Their corner plots also show a degeneracy between the abundance, cloud-top pressure, and temperature which could explain the discrepancy. The consistency of this degeneracy, regardless of analysis framework, suggests a need for analyses covering a broader wavelength range that can better constrain cloud properties. 

Constraints on the atmospheric CO abundance of HD 189733 b have been more elusive. \citet{dekok2013} reported the first CO detection for HD 189733 b, also based on $K$-band emission spectroscopy, but did not report an abundance due to uncertainties concerning the impact of hazes. \citet{brogi2019} re-analyzed the \citet{dekok2013} observations of HD 189733 b to retrieve CO and H$_2$O, but obtained only lower and upper limits, respectively. The analysis presented in Section \ref{sec:res} is the first simultaneous determination of CO, CH$_4$, and H$_2$O abundances in the atmosphere of HD 189733 b suitable for constraining the atmospheric C/O ratio. Several \textit{JWST} programs have obtained spectroscopy of HD 189733 b in both near and mid infrared wavelengths (GO 1633, PI Deming; GO 2001, PI Min; GO 2021, PI Kilpatrick \& Kataria), which should provide a second, independent estimate of the atmospheric metallicity and C/O ratio (including the impact of CO$_2$) to compare with our results.

\subsection{Winds and circulation}

Compared with the nominal values in Table \ref{tab:props}, the retrieved $\Delta K_{\rm p}$ and $\Delta v_{\rm sys}$ yield a blueshift of 8 \kms at the start of the observation sequence, increasing to 10 \kms by the end. While wind speeds as high as 8 \kms have been reported for HD 189733 b from transmission observations of sodium lines \citep{wytenbach2015}, other observations in the near-infrared have preferred a windspeed of 2 \kms \citep{brogi2016}, while still other infrared transmission analyses have reported $K_{\rm p}$ and $v_{\rm sys}$ values compatible with those reported in Table \ref{tab:priors} \citep{Boucher2021, klein2023}. We expect that a day-to-night wind would appear as a redshift in post-eclipse emission observations, rather than a blueshift. Similarly, we would expect the hotter dayside dominating the overall emission signal of the rotating planet to lead to a slight additional redshift. Improved orbital phase coverage may be required to directly constrain the wind speeds and 3D circulation on HD 189733 b from high-resolution spectroscopy alone.

Factors other than planetary winds could be contributing to the observed blueshift. The velocity shift is too large to be easily explained by an error in the orbital phase, particularly given the precision of the transit time for HD 189733 b. Similarly, we expect our wavelength solution to be reliable to $\sim 100 \rm\,m\ s^{-1}$, substantially better than the observed shift. \citet{rasmussen2023} demonstrated that the shift in planetary lines within a science exposure can produce a blueshift of several \kms, though we expect this effect to be minimal given the $<1$\,\kms velocity shift between successive spectra in the time series. However, we did not attempt to account for the wavelength-dependence or non-Gaussian wings of the NIRSPEC line-spread function \citep{nirspecupgrade2}, potentially leading to a model mismatch in the line shapes that could bias the retrieved velocities. A laser frequency comb compatible with KPIC is scheduled for deployment in late 2023, which, when completed, should enable the more robust calibration of the line-spread function required to interpret these types of velocity offsets as physical wind speeds or 3D effects. We note that a similar blueshift was also reported in \citet{finnerty2023} for WASP-33 b, which may indicate a systematic effect in the retrieval pipeline.

While we included a rotational broadening kernel in our fit, our approach to the instrumental broadening makes this effectively a nuisance parameter rather than a reliable estimate of the planetary $v\sin i$. As discussed above, we ignore both the wavelength-dependence and non-Gaussian shape of the line-spread function, as well as the impact of line smearing from the change in planet velocity within an exposure. To account for this, we chose a slightly smaller than expected Gaussian kernel for the instrument profile, and allowed a larger rotational kernel to make up the difference based on the actual observed line widths after coadding fibers. Future improvements to our retrieval pipeline will include a wavelength-dependent instrument profile similar to that used by \citet{wang2021} to measure $v\sin i$ for brown dwarf companions and directly-imaged planets observed with KPIC. Combined with the use of the laser frequency comb to constrain relative changes in the LSF across the NIRSPEC detector, this should enable robust measurements of $v\sin i$ for hot Jupiters in the future with KPIC.

\subsection{CH$_4$ and photochemistry}

\begin{figure}
    \centering
    \includegraphics[width=1.0\linewidth]{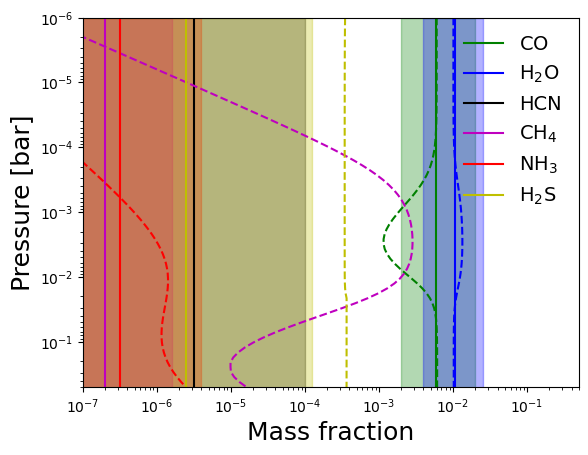}
    \caption{Retrieved molecular abundances (solid lines, 1$\sigma$ shaded) compared with the equilibrium abundances estimated from \texttt{easychem} (dashed lines). The equilibrium abundances are consistent with $\rm C/O = 0.25$ and $\rm [M/H] = +0.05$, both slightly below the values obtained from the gas-phase-only retrieval. The equilibrium model predicts H$_2$S and CH$_4$ abundances substantially greater than the retrieved upper limits. These species may be photochemically depleted. }
    \label{fig:chemplot}
\end{figure}

Similar to \citet{finnerty2023}, we compare our retrieved abundances with the results of the equilibrium chemistry calculator \texttt{easyCHEM} \citep[][Lei et al. in prep.]{molliere2017} for the retrieved $P-T$ profile. Figure \ref{fig:chemplot} plots the equilibrium vertical abundance profiles with the constant vertical abundances from our retrievals. The retrieved CO and H$_2$O abundances are a good match to the equilibrium predictions for C/O = 0.25 and [M/H] = 0.05, suggesting the gas-phase retrieval may be slightly overestimating the atmospheric metallicity.  Both H$_2$O and CO abundances are roughly constant with height in equilibrium, validating the constant-with-altitude assumption in our retrieval and indicating that the potential biases resulting from high-altitude H$_2$O depletion in ultra-hot Jupiters \citep{brogi2023} are not a factor for HD 189733 b.

The cumulative distribution function (CDF) of the retrieved CH$_4$ posterior indicates $\rm \log CH_4 < -4.6$ with 99\% confidence. However, Figure \ref{fig:chemplot} indicates that in chemical equilibrium for the retrieved $P-T$ profile, the CH$_4$ mass fraction should be $>10^{-4}$ over nearly the entire pressure range probed by the KPIC observations in order to match the retrieved H$_2$O and CO abundances. We caution that the retrieved $P-T$ profile in Figure \ref{fig:spec} is somewhat colder than expected in the upper atmosphere, which may lead to an overprediction of CH$_4$ when assuming chemical equilibrium. While previous attempts at CH$_4$ detection in hot Jupiters have faced challenges with linelist accuracy, we use an opacity table based on the recent \citet{hargreaves2020} linelist which has been validated on brown dwarfs of similar effective temperature \citep{tannock2022}. This opacity table has also previously been used to successfully measure the CH$_4$ abundance in a brown dwarf companion from KPIC observations at a mixing ratio comparable to that expected in HD 189733 b \citep{xuan2022}. This suggests that the retrieved upper limit on CH$_4$ is due to a real absence of the expected CH$_4$, rather than a linelist issue.

Depletion of CH$_4$ could be due to photochemistry. We assessed this possibility using the \texttt{VULCAN} chemical kinetics code \citep{vulcan}. We used the default $P-T$/$K_{zz}$ profile for HD 189733 b included with \texttt{VULCAN} and described in \citet{vulcan}, which is slightly hotter than our retrieved profile in the upper atmosphere, the SNCHO photochemistry network, and the C/H and O/H set to the retrieved medians. All other settings were unchanged from the \texttt{VULCAN} defaults. \texttt{VULCAN} predicts a roughly constant CH$_4$ mass fraction of $\sim10^{-5}$ from at pressures between approximately 1 and $10^{-3}$ bar. Over the same pressure range, the NH$_3$ mass fraction is predicted to be $\sim10^{-4}$. The abundance of both species rapidly drops above 1 mbar, which may bias the abundances obtained from a free retrieval assuming constant abundance with pressure towards lower values, similar to the impact of  water depletion in the upper atmospheres of ultra-hot Jupiters discussed in \citet{brogi2023}.

From the retrieved posterior, we obtain 99\% upper limits $\rm \log CH_4 < -4.6$ and $\rm \log NH_3 < -4.5$, suggesting the current KPIC observations are not quite sensitive enough to make a definitive (non)detection at the \texttt{VULCAN}-predicted abundances. Both CH$_4$ and NH$_3$ have significant opacity throughout the NIR. Incorporating additional data such as high-resolution $L$-band observations from KPIC phase III may improve sensitivity to these species and enable tests of photochemical models in the future. Additionally, incorporating flux-calibrated broad-band medium-resolution observations from \texttt{JWST} would significantly improve constraints on absolute abundances compared with high-resolution observations alone, and also enable constraints on species which lack features easy to observe from the ground, such as CO$_2$

\subsection{$^{13}$CO enrichment}\label{sec:coenrich}

While the CO isotopologue ratio is not well-constrained in the retrieval, comparing the log-likelihood of the maximum-likelihood model with and without $\rm ^{13}CO$ indicates the presence of $\rm ^{13}CO$ is favored at $\sim3\sigma$. The marginalized posterior shows a clear preference for high levels of $\rm ^{13}CO$ enrichment compared with both the solar system value of $\rm ^{12}C/^{13}C\sim89$ and local interstellar value of $\rm ^{12}C/^{13}C\sim68$ \citep{woods2009, milam2005}, but with a long tail that is still consistent with the interstellar value at $1\sigma$. Previously, \citet{finnerty2023} and \citet{line2021} reported indications of $\rm ^{13}CO$ enrichment in hot Jupiter atmospheres compared with the local ISM. These estimates are substantially less than the $\rm ^{12}CO/^{13}CO\sim10^{+53}_{-6}$ median retrieved for HD 189733 b, and were instead broadly consistent with the level of isotopic fractionation expected in the midplanes of protoplanetary disks \citep{woods2009}. 

Carbon isotopologue constraints have also been obtained for widely separated companions. \citet{zhang2021} found a roughly solar value for the $\rm ^{12}CO/^{13}CO$ ratio in a young field brown dwarf and suggested a formation process with minimal preferential accretion of $\rm ^{13}CO$-enriched material such as direct gravitational collapse. In contrast, $\rm ^{13}CO$ enrichment has been reported for a young accreting super-Jupiter \citep{zhang2021enriched}, possibly suggesting preferential accretion of $\rm ^{13}C$ via fractionated ices. The enrichment reported in \citet{zhang2021enriched} is similar to the values reported in \citet{line2021} for WASP-77Ab and \citet{finnerty2023} for WASP-33b, still less than the preferred value of $\sim10$ for HD 189733 b. 

The potential prevalence of $\rm ^{13}CO$ enrichment in hot Jupiters observed to-date suggests a common process in the formation or evolution of these planets. Accretion of fractionated ices appears to be consistent with the values reported by \citet{line2021} and \citet{finnerty2023} \citep{woods2009}, but may not be sufficient to explain the $\rm ^{12}CO/^{13}CO\sim10$ preference seen in our HD 189733 b retrieval. Additional observational and modeling work is required to constrain the $\rm ^{12}CO/^{13}CO$ ratios of host stars and understand the general prevalence of this $^{13}$CO enrichment, and the potential extreme case of HD 189733 b in particular.

\subsection{C/O and metallicity}

\begin{figure}
    \centering
    \includegraphics[width=1.0\linewidth]{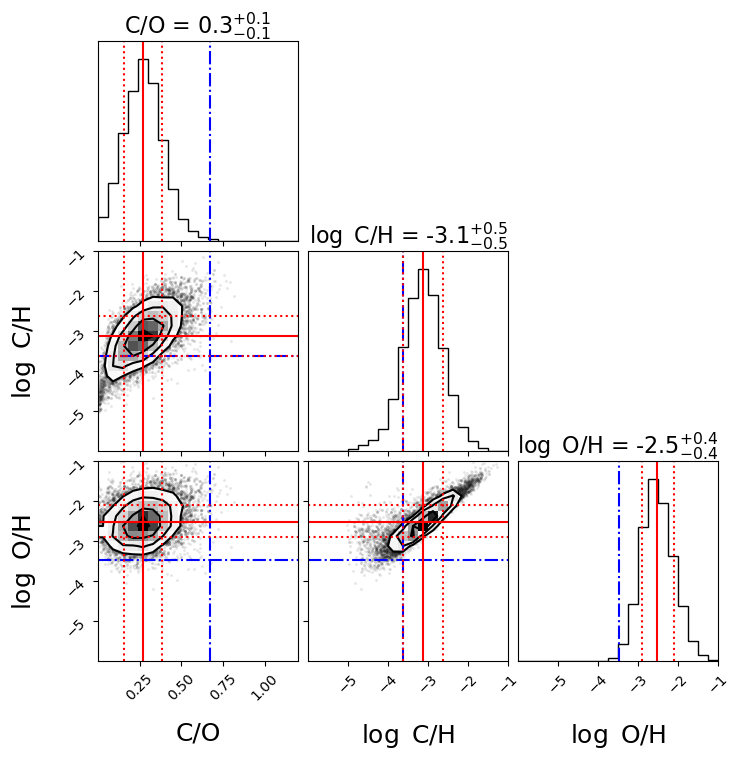}
    \caption{Retrieved C/O, C/H, and O/H posteriors. The median value is indicated in solid red, with the upper and lower bounds of the 68\% confidence interval in dotted red. Stellar values from \citet{polanski2022} are shown in dash-dot blue. The retrieved C/O ratio is substantially sub-stellar, while the carbon and oxygen abundances are both super-stellar. }
    \label{fig:CO}
\end{figure}

The posteriors for C/O, C/H, and O/H derived from the retrieval are plotted in Figure \ref{fig:CO}. As in previous high-resolution studies \citep[e.g.][]{xuan2022, finnerty2023}, the ratio of species is better constrained than the absolute abundances, giving $\rm C/O = 0.3\pm0.1$, $\rm \log C/H = -3.1\pm0.5$, and $\rm \log O/H = -2.5\pm0.4$. Precise abundances for HD 189733 A were previously reported in \citet{polanski2022}, enabling a robust comparison between the composition of the star and the planetary atmosphere. The CDFs of the distributions in Figure \ref{fig:CO} give a sub-stellar C/O at 99.8\% confidence, super-stellar C/H at 84\% confidence, and super-stellar O/H at 99.4\% confidence. Comparison to equilibrium models suggests the gas-only retrieval may be slightly overestimating the total atmospheric metallicity, but sub-solar metallicities are still disfavored. These results indicate the atmosphere of HD 189733 b is compositionally distinct from the host star.

The combination of substellar-C/O and super-stellar metallicity was suggested to be an indicator of significant post-formation ice accretion by \citet{oberg2011}. \citet{madhusudhan2014} found that low-C/O, metal-enriched atmospheres could be indicative of either core accretion followed by disk migration or formation at wide separations via gravitational instability followed by disk-free migration. Subsequently, \citet{madhusudhan2017} found that such an atmospheric composition could also be the result of core erosion or late planetesimal accretion. 

Core accretion should lead to atmospheric metal enrichment in giant planets \citep{thorngren2016}, including a trend between planet mass and metal enrichment relative to the host star that is consistent with our reported values for HD 189733 b \citep{cridland2019}. This process is also predicted to produce an inverse correlation between atmospheric metallicity and C/O \citep{espinoza2017, cridland2019, khorshid2022}, though this trend may not hold in the presence of pebble drift \citep{booth2017}. These models can predict the observed composition of HD 189733 b as a result of accreting a significant ($\sim10\%$) fraction of the total planetary mass as solids relatively late in the formation process, with radial pebble drift not substantially altering the overall C/O of this material. Such a scenario may also be consistent with the extremely high $^{13}$CO enrichment suggested by the retrieval, which could be explained through late accretion of a large amount of highly fractionated ice. 

\section{Summary and Conclusions}\label{sec:conc}

We performed a Bayesian retrieval on high-resolution $K$-band observations of HD 189733 b from Keck/KPIC, successfully constraining the abundances of CO and H$_2$O and placing upper limits on the atmospheric mass fractions of CH$_4$, NH$_3$, and HCN. The retrieved abundances yield an atmospheric $\rm C/O = 0.3\pm0.1$ for HD 189733 b, substantially less than the stellar C/O ratio, and a stellar-to-superstellar atmospheric metallicity. This composition could be explained by the accretion of a large amount of solid material late in the planet formation process, possibly as a result of disk migration, which could also produce a well-aligned orbit with respect to the stellar rotation. 

We do not detect CH$_4$ in the atmosphere of HD 189733 b. While this is incompatible with the retrieved $P-T$ profile in chemical equilibrium, the upper limits we obtained may be consistent with models which incorporate atmospheric photochemistry. Additional observations with a wider wavelength coverage will improve sensitivity to both CH$_4$ and NH$_3$, enabling direct tests of photochemical models, haze formation mechanisms, and measurement of the N/O ratio -- providing deeper insight into the formation and evolution of the HD 189733 planetary system.

\begin{acknowledgments}
We thank the anonymous referee whose detailed and insightful comments improved the quality of this paper. L. F. is a member of UAW local 2865. L.F. acknowledges the support of the W.M. Keck Foundation, which also supports development of the KPIC facility data reduction pipeline. The contributed Hoffman2 computing node used for this work was supported by the Heising-Simons Foundation grant \#2020-1821. Funding for KPIC has been provided by the California Institute of Technology, the Jet Propulsion Laboratory, the Heising-Simons Foundation (grants \#2015-129, \#2017-318, \#2019-1312, \#2023-4597, \#2023-4598), the Simons Foundation (through the Caltech Center for Comparative Planetary Evolution), and the NSF under grant AST-1611623. D.E. is supported by a NASA Future Investigators in NASA Earth and Space Science and Technology (FINESST) fellowship under award \#80NSSC19K1423. D.E. also acknowledges support from the Keck Visiting Scholars Program (KVSP) to install the Phase II upgrades required for KPIC VFN.

This work used computational and storage services associated with the Hoffman2 Shared Cluster provided by UCLA Institute for Digital Research and Education’s Research Technology Group. L.F. thanks Briley Lewis for her helpful guide to using Hoffman2, and Paul Molli\`ere for his assistance in adding additional opacities to petitRADTRANS. 

The data presented herein were obtained at the W. M. Keck Observatory, which is operated as a scientific partnership among the California Institute of Technology, the University of California and the National Aeronautics and Space Administration. The Observatory was made possible by the generous financial support of the W. M. Keck Foundation. The authors wish to recognize and acknowledge the very significant cultural role and reverence that the summit of Mauna Kea has always had within the indigenous Hawaiian community.  We are most fortunate to have the opportunity to conduct observations from this mountain. 

This research has made use of the NASA Exoplanet Archive, which is operated by the California Institute of Technology, under contract with the National Aeronautics and Space Administration under the Exoplanet Exploration Program.

\end{acknowledgments}

%

\vspace{5mm}
\facilities{Keck:II(NIRSPEC/KPIC)}


\software{astropy \citep{astropy:2013, astropy:2018},  
          \dynesty\ \citep{speagle2020},
          \texttt{corner} \citep{corner}, \texttt{VULCAN} \citep{vulcan}
          \petit\ \citep{prt:2019, prt:2020}}


\appendix
\section{Corner Plots}\label{app:corner}
\begin{figure}
    \centering
    \includegraphics[width=1.0\linewidth]{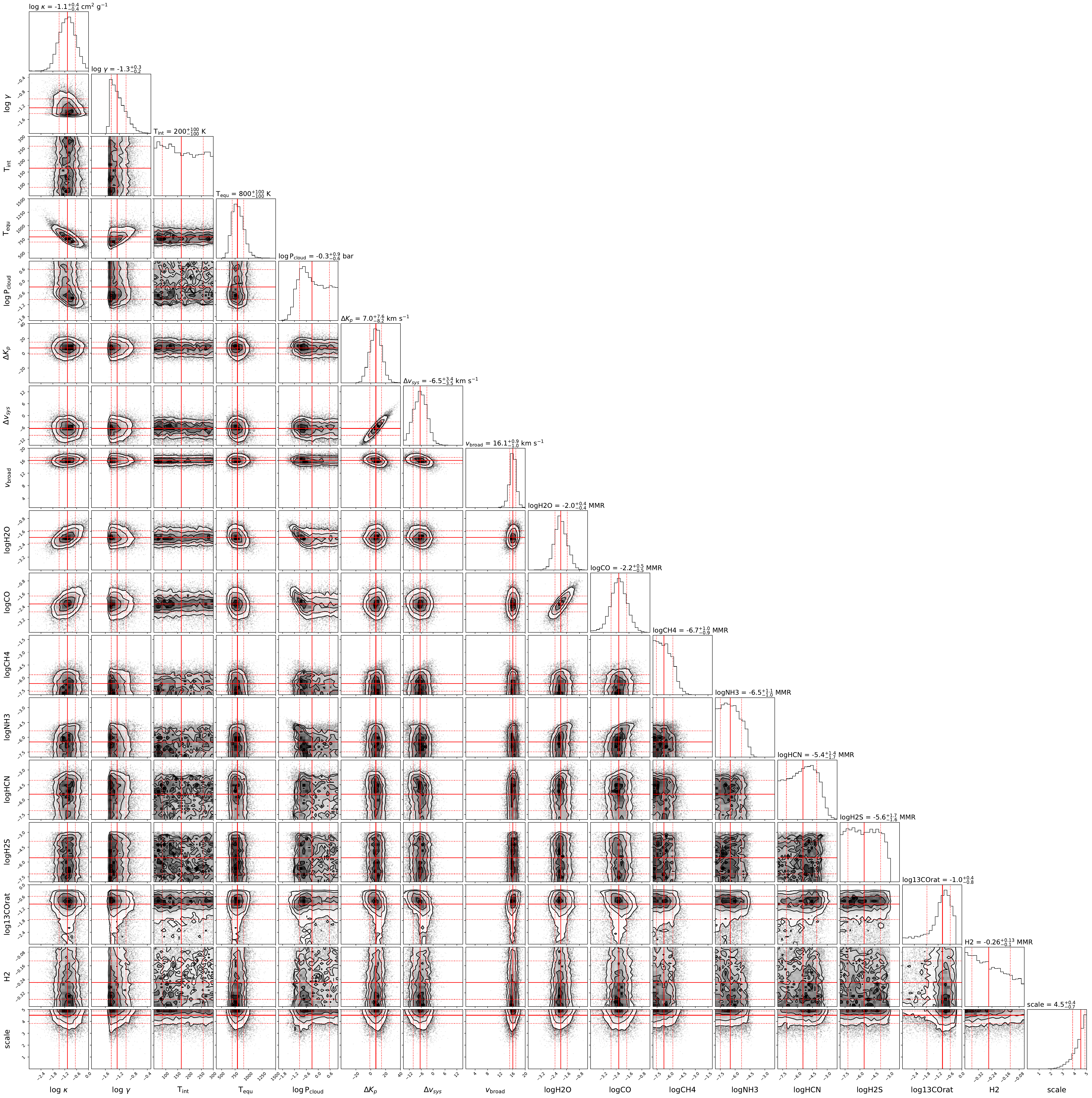}
    \caption{Full corner plot for the retrieval presented in Section \ref{sec:res}. Red solid lines indicate the medians, while red dashed lines indicate the bounds of the marginalized 68\% confidence interval. We discuss these results in Section \ref{sec:res}.}
    \label{fig:corner}
\end{figure}

Figure \ref{fig:corner} presents the full corner plot from the retrieval for completeness. We discuss the retrievals in Section \ref{sec:res}, including poorly constrained and degenerate parameters. The units, priors, and retrieved quantity are listed in Table \ref{tab:priors}

\section{Molecular detection maps}\label{app:detmaps}

\begin{figure}
    \centering
    \includegraphics[width=1.0\linewidth]{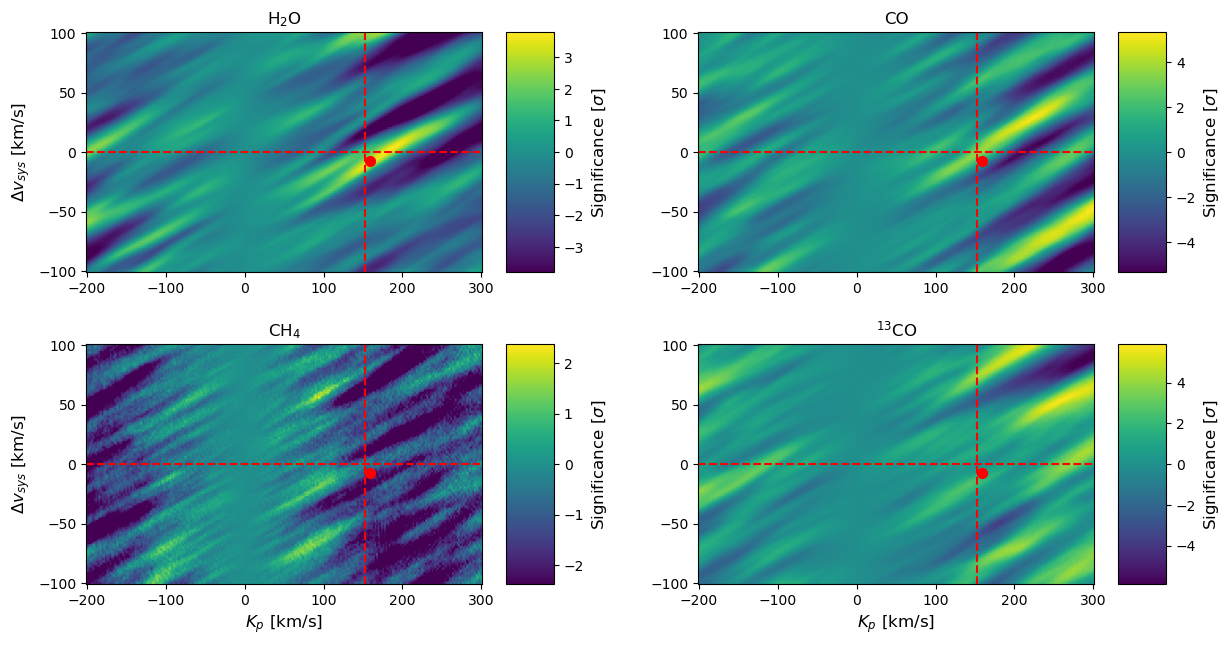}
    \caption{$v_{\rm sys} - K_{\rm p}$ diagrams for H$_2$O, CO, CH$_4$, and $\rm^{13}CO$. Dashed red lines indicate the nominal values of the planetary velocity parameters, and the red dot indicates the maximum-likelihood values for the full model. These maps were made by subtracting a map omitting each species from a map made using the maximum-likelihood model, in order to better account for the impact of other atmospheric species on the continuum. Significance was estimated by dividing each map by the standard deviation of the $K_{\rm p} < 0$ region. Continuum impacts are still significant, leading to weaker-than-expected detections of H$_2$O and CO and biasing the CO velocity. These effects are less important for lower-abundance species. As expected based on the retrieved posteriors, we do not detect CH$_4$, and $\rm^{13}CO$ shows a weak ($\sim3\sigma$) feature near the expected planet velocity. This $\rm^{13}CO$ feature does not constitute and independent detection, but the inclusion of $\rm^{13}CO$ is favored over a $\rm^{12}CO$-only model at $\sim3\sigma$, consistent with other results for $\rm^{13}CO$ in hot Jupiter atmospheres \citep{line2021, finnerty2023}}
    \label{fig:detmaps}
\end{figure}

Figure \ref{fig:detmaps} presents $v_{\rm sys} - K_{\rm p}$ diagrams for each detected species calculated using a leave-one-out approach. As shown in Figure \ref{fig:spec}, strong, broad absorption by CO and H$_2$O cause significant changes to the overall shape of the planetary SED, particularly beyond $\rm \sim2.35\mu m$. This presents a challenge when using single-molecule templates to assess detection strength, as the resulting inaccuracies in the continuum level can make it impossible for a single-molecule template to reproduce the observed relative fluxes between lines, particularly if other parameters are held fixed.

A slightly more reliable approach is to calculate $v_{\rm sys} - K_{\rm p}$ diagram for the maximum-likelihood model as well as the $v_{\rm sys} - K_{\rm p}$ diagram for a model omitting the molecule of interest. The difference between these two maps is the change in the log-likelihood that can be attributed to the presence of the omitted species. This approach will still be inaccurate in the case of e.g. H$_2$O and CO whose omission has a significant impact on the continuum, but should provide a more accurate estimate for the detection strength of trace species. Taking the difference of the two maps should also largely remove the self-division artifact seen at $K_{\rm p} = 0$ in Figure \ref{fig:kpvsys}, though the resulting map will show a bias towards higher values of $K_{\rm p}$. This bias arises from self-removal of the planet model during detrending and is increased by the explicit dependence of the \citet{brogi2019} log-likelihood function on the model variance. The detrending process used to remove continuum and telluric features reduces the model variance as $K_{\rm p}$ decreases and the self-division effect of the median division and the self-subtraction effect of the SVD remove more of the planet model. As a result, we expect that the apparent strength of noise features to increase with increasing $K_{\rm p}$ when assuming a constant variance for the entire map, and that planetary features will show a bias towards higher $K_{\rm p}$ in difference maps compared with computing $\log L$ directly.

These effects can be seen in Figure \ref{fig:detmaps}. H$_2$O is detected at $\sim3.8\sigma$ near the expected velocity, while CO is detected at $\sim5\sigma$ but with a shift towards larger $K_{\rm p}$. CH$_4$ is not significantly detected, as expected, and a weak ($\sim3\sigma$) feature is present in the $\rm^{13}CO$ map near the maximum-likelihood velocities, but is not strong enough to constitute an independent detection, particularly given the presence of large-amplitude features at high $K_{\rm p}$. The H$_2$O and CO detections are weaker than would be expected based off of the all-molecule model, likely due to the impact of these species on the planet continuum. However, $\rm^{13}CO$ is significantly less abundant and will have a much smaller impact on the planetary continuum level. Consistent with this, the $\rm^{13}CO$ feature in the $v_{\rm sys} - K_{\rm p}$ map is similar in strength to the expectation from Wilks' Theorem in Section \ref{sec:coenrich}.

Consistent with previous high-resolutions studies of other hot Jupiters \citep{line2021,finnerty2023}, $\rm^{13}CO$ is not independently detected in the $v_{\rm sys} - K_{\rm p}$ space, but the presence of $\rm^{13}CO$ is favored by $\sim3\sigma$ over a model without $\rm^{13}CO$. As discussed above, estimating the detection strength from the $v_{\rm sys} - K_{\rm p}$ diagram poses a number of challenges that are particularly acute for trace species which do not produce a strong peak to begin with. The repeated tentative detection of $\rm^{13}CO$ across multiple targets highlights the need for improved techniques to quantitatively estimate detection strengths for high-resolution observations.

\clearpage
\bibliography{exoplanetbib}{}
\bibliographystyle{aasjournal}



\end{CJK*}
\end{document}